\documentclass[longauth]{aa}  
\usepackage{graphicx}
\usepackage{txfonts}
\usepackage{xcolor}
\usepackage{amsmath}

\usepackage[hidelinks,colorlinks=true,linkcolor=blue,anchorcolor=black,citecolor=blue,filecolor=black,menucolor=black,runcolor=black,urlcolor=black]{hyperref}
\usepackage{orcidlink}
\usepackage{ragged2e}
\usepackage{float}
\usepackage{rotating}
\usepackage{multirow}
\usepackage{placeins}
\usepackage{array}
\usepackage{lipsum}
\usepackage{afterpage}
\usepackage[most]{tcolorbox}

\def\Msun{M$_\odot$}

\def\chaha1{Cha~H$\alpha$~1}
\newcommand{\mum}{$\mu$m\xspace}

\begin{document}

   \title{MINDS. Cha H$\alpha$ 1, a brown dwarf with a hydrocarbon-rich disk.}

   \author{Mar\'ia Morales-Calder\'on \inst{\ref{inst1}}
          \and
           Hyerin Jang \inst{\ref{inst2}}          
          \and
          Aditya M. Arabhavi  \inst{\ref{inst3}}
          \and    
         Valentin Christiaens \inst{\ref{inst4}, \ref{inst5}}
         \and    
         David Barrado\inst{\ref{inst1}}
         \and    
	Inga Kamp\inst{\ref{inst3}}
         \and    
	Ewine F. van Dishoeck\inst{\ref{inst6}, \ref{inst7}}
         \and    
	Thomas Henning\inst{\ref{inst8}}
         \and    
	L. B. F. M. Waters\inst{\ref{inst2}}
         \and    
	Milou Temmink\inst{\ref{inst6}}
         \and    
	Manuel G\"udel\inst{\ref{inst9}, \ref{inst10}}
         \and    
	Pierre-Olivier Lagage\inst{\ref{inst11}}
         \and    
	Alessio Caratti o Garatti\inst{\ref{inst12}}
         \and    
	Adrian M. Glauser\inst{\ref{inst10}}
         \and    
	Tom P. Ray\inst{\ref{inst13}}
         \and    
	Riccardo Franceschi\inst{\ref{inst14}}
        \and    
	Danny Gasman\inst{\ref{inst4}}
         \and    
	Sierra L. Grant\inst{\ref{inst15}}
         \and    
	Till Kaeufer\inst{\ref{inst16}}
         \and    
	Jayatee Kanwar\inst{\ref{inst3},\ref{inst17},\ref{inst18}}
         \and    
	Giulia Perotti\inst{\ref{inst19},\ref{inst8}}
         \and    
	Matthias Samland\inst{\ref{inst8}}
         \and    
	Kamber Schwarz\inst{\ref{inst8}}
         \and    
	Marissa Vlasblom\inst{\ref{inst6}}
         \and    
	Luis Colina\inst{\ref{inst1}}
         \and    
	G\"oran \"Ostlin\inst{\ref{inst20}}
          }

\institute{Centro de Astrobiolog\'ia (CAB), CSIC-INTA, ESAC Campus, Camino Bajo del Castillo s/n, 28692 Villanueva de la Ca\~nada, Madrid, Spain\label{inst1}\\
             \email{mariamc@cab.inta-csic.es}
         \and
         Department of Astrophysics/IMAPP, Radboud University, PO Box 9010, 6500 GL Nijmegen, The Netherlands\label{inst2}
         \and
         Kapteyn Astronomical Institute, Rijksuniversiteit Groningen, Postbus 800, 9700AV Groningen, The Netherlands \label{inst3}
         \and
         Institute of Astronomy, KU Leuven, Celestijnenlaan 200D, 3001 Leuven, Belgium \label{inst4}
         \and
         STAR Institute, Universit\'e de Li\`ege, All\'ee du Six Ao\^ut 19c, 4000 Li\`ege, Belgium \label{inst5}
         \and
	Leiden Observatory, Leiden University, PO Box 9513, 2300 RA Leiden, the Netherlands \label{inst6}
         \and
         Max-Planck Institut f\"{u}r Extraterrestrische Physik (MPE), Giessenbachstr. 1, 85748, Garching, Germany\label{inst7}
         \and
	Max-Planck-Institut f\"{u}r Astronomie (MPIA), K\"{o}nigstuhl 17, 69117 Heidelberg, Germany \label{inst8}
         \and
	Dept. of Astrophysics, University of Vienna, T\"urkenschanzstr. 17, A-1180 Vienna, Austria \label{inst9}
         \and
	ETH Z\"urich, Institute for Particle Physics and Astrophysics, Wolfgang-Pauli-Str. 27, 8093 Z\"urich, Switzerland \label{inst10}
         \and
	Universit\'e Paris-Saclay, Universit\'e Paris Cit\'e, CEA, CNRS, AIM, F-91191 Gif-sur-Yvette, France \label{inst11}
         \and
	INAF – Osservatorio Astronomico di Capodimonte, Salita Moiariello 16, 80131 Napoli, Italy \label{inst12}
         \and
	Dublin Institute for Advanced Studies, 31 Fitzwilliam Place, D02 XF86 Dublin, Ireland \label{inst13}
         \and
	LESIA, Observatoire de Paris, Universit\'e PSL, CNRS, Sorbonne Universit\'e, Universit\'e de Paris, 5 place Jules Janssen, 92195 Meudon, France \label{inst14}
         \and
	Earth and Planets Laboratory, Carnegie Institution for Science, 5241 Broad Branch Road, NW, Washington, DC 20015, USA\label{inst15}
         \and
	Department of Physics and Astronomy, University of Exeter, Exeter EX4 4QL, UK\label{inst16}
	\and
	Space Research Institute, Austrian Academy of Sciences, Schmiedlstr. 6, A-8042, Graz, Austria \label{inst17}
         \and
	TU Graz, Fakultät für Mathematik, Physik und Geodäsie, Petersgasse 16 8010 Graz, Austria \label{inst18}
         \and
	Niels Bohr Institute, University of Copenhagen, NBB BA2, Jagtvej 155A, 2200 Copenhagen, Denmark \label{inst19}
	\and
	Department of Astronomy, Oskar Klein Centre; Stockholm University; SE-106 91 Stockholm, Sweden  \label{inst20}
             }


  \abstract
   {The chemistry of disks around brown dwarfs (BDs) remain largely unexplored due to their faintness. Despite the efforts performed with Spitzer, we have far less understanding of planet formation, chemical composition, disk structure, and evolution in disks around BDs compared to their more massive counterparts (T Tauri and Herbig Ae/Be stars), which are more readily studied due to their greater brightness. Recent JWST observations, with up to an order of magnitude improvement in both spectral and spatial resolution, have shown that these systems are chemically rich, offering valuable insights into giant planet formation.}
{As part of the MIRI mid-INfrared Disk Survey (MINDS) JWST guaranteed time program, we aim to characterize the gas and dust composition of the disk around the brown dwarf [NC98]~Cha~HA~1, hereafter \chaha1, in the mid-infrared.}
   {We obtain data from the MIRI Medium Resolution Spectrometer (MRS) from 4.9 to 28~\mum ($R\sim$1500 - 3500; FWHM$\sim$0.2" - 1.2"). We use the dust fitting tool DuCK to investigate the dust composition and grain sizes while we identify and fit molecular emission in the spectrum using slab models.}
   {Compared with disks around very low mass stars, clear silicate emission features are seen in this BD disk. In addition, JWST reveals a plethora of hydrocarbons, including C$_2$H$_2$, $^{13}$CCH$_2$, CH$_3$, CH$_4$, C$_2$H$_4$, C$_4$H$_2$,  C$_3$H$_4$, C$_2$H$_6$,  and C$_6$H$_6$ which suggest a disk with a gas C/O > 1. Additionally, we detect CO$_2$, $^{13}$CO$_2$, HCN, H$_2$, and H$_2$O. CO and OH are absent from the spectrum.
The dust is dominated by large $\sim$4~\mum  size amorphous silicates (MgSiO$_3$). We infer a small dust mass fraction ($>$10$\%$) of 5~\mum  size crystalline forsterite. We do not detect polycyclic aromatic hydrocarbons. }
{The mid-infrared spectrum of  \chaha1 shows the most diverse chemistry seen to date in a BD protoplanetary disk, consisting of a strong dust feature, 12 carbon-bearing molecules plus H$_2$, and water. The diverse molecular environment offers a unique opportunity to test our understanding of BD disks chemistry and how it affects the possible planets forming in them.}

   \keywords{(Stars:) brown dwarfs -- Protoplanetary disks -- Infrared: planetary systems --
               }

   \maketitle

\section{Introduction}

Brown dwarfs (BDs) are often referred to as "failed stars" because they lack sufficient mass ($\le$0.075 M$_\odot$ for solar metallicity) to sustain the nuclear fusion of hydrogen into helium in their cores. The presence of warm dusty disks around these dim objects is well known \citep{Comeron98, Natta01, Natta02}, based on near- and mid-infrared studies.  These disks can result in the formation of Earth-sized or smaller planets \citep{Apai05, Payne07, Daemgen16}.  \cite{Chauvin04} were the first to report a giant planetary candidate around a brown dwarf. In the following years, other brown dwarfs with Jupiter-mass planets have been detected  \cite[e.g., ][]{Todorov10, Han13, Jung18, Barrado07b}. 

The study of the properties of disks around young brown dwarfs provide important clues on the formation of the planetary systems forming around them. Though usually smaller in size and mass \citep{Hendler17}, BD protoplanetary disks are similar to the disks around their young stellar counterparts (e.g., T Tauri stars, TTS) in many ways. Both BD disks and TTS disks are typically flat \citep{Scholz07} with evidence of dust grain growth in their inner regions \citep{Sterzik04, Apai05, Meru13, Pinilla13}. Some BD disks also show signs of inner disk dust clearing \citep{Muzerolle06, Rilinger19}. Furthermore, the fraction of BDs observed to host disks is similar to the TTS disk fraction, possibly indicating a shared formation mechanism \citep{Monin10} and outflows have been observed in these faint objects \citep{Whelan07}, which indicate similar accretion/ejection mechanisms in BDs as in their higher mass counterparts. However, some aspects of disks around BDs are different from those around T Tauri and Herbig Ae/Be stars (intermediate mass pre-main sequence stars, 2 to 8~M$_\odot$, the higher mass analogs to the T Tauri stars). They have a lower accretion rate  \cite[$\sim$10$^{-12}$~M$_\odot$yr$^{-1}$;][]{Herczeg09}, a tendency to show flatter disk geometry inferred from Spitzer Space Telescope observations and spectral energy distribution (SED) modeling \cite[e.g., ][]{Apai05, Guieu07, Scholz07, Pascucci09}, and a longer lifetime \citep{Barrado03, Carpenter06, Barrado07}. The dust in BD disks seems to undergo more efficient radial drift than in higher mass systems \citep{Pinilla13}. The disk chemistry and physical evolution also show important differences as a function of stellar mass, temperature, and radiation field \cite[e.g., different dust populations at the observed radii or different molecular abundances in the disk atmospheres;][]{Pascucci09}. This is important for planet formation models since the different chemistry of materials in the disk may also influence the bulk composition and volatile content of the forming planets. 

Thanks  to Spitzer, the Herschel Space Observatory, and the Atacama Large Millimeter Array, the past few years have seen a dramatic progress in understanding disks around the lowest mass stars and brown dwarfs \cite[e.g., ][]{Pascucci13, Riviere15, Testi16, Kurtovic20}.  However, our understanding of planet formation, chemical composition, disk structure, and evolution around BDs lags behind that of their higher mass counterparts, such as T Tauri and Herbig Ae/Be stars. The latter are easier to characterize due to their brighter nature  and have been studied in larger numbers. 
The James Webb Space Telescope (JWST) sensitivity enables the characterization of the disks around BDs in a wavelength range previously inaccessible, revealing insights into physical processes of the gas and dust. Very few BD disks have been observed to date with the JWST. \cite{Perotti25} presented the Mid-InfraRed Instrument \cite[MIRI;][]{Rieke15,Wright15,Wright23} spectrum of a water rich, highly inclined BD disk which appears to be poor in hydrocarbons. On the contrary, TWA 27 A has a disk with no water or silicate emission but rich in hydrocarbons \citep{Patapis25}. Similarly, recent JWST observations of Very Low-Mass Stars (VLMS; M $<$ 0.3 M$_{\sun}$) have shed new light on the chemical composition of their disks revealing a complex chemistry rich in hydrocarbons, including some isotopologues  \cite[][]{Tabone23, Arabhavi24, Kanwar24}. These disks around VLMS, including the disk around TWA 27 A, display strong hydrocarbon emissions (in some cases even a hydrocarbon pseudo-continuum) with very few oxygen carriers suggesting a high volatile carbon-to-oxygen (C/O) ratio. 
A larger sample of objects observed, and extension of the sample to lower masses, will inform us on which is the norm among BD disks (see \cite{Arabhavi25b} for a study on trends in the inner disk gas of VLMS disks) and will help us investigate the various conditions under which planets may form.

The Cycle 1 JWST/MIRI Guaranteed Time Observation program MINDS (MIRI mid-INfrared Disk Survey\footnote{https://minds.cab.inta-csic.es/}), aims at investigating the chemical inventory in the terrestrial planet-forming zone of disks from Herbig Ae/Be stars to BDs \citep{Henning24,Kamp23}. As part of this program, we present here JWST spectroscopic observations of the disk around the brown dwarf [NC98]~Cha~HA~1 (hereafter \chaha1) which has been characterized in detail in \cite{Bayo17}. \chaha1 is a member of the roughly 1.4 - 2.4~Myr-old Chamaeleon I South star forming region at a distance of 191.8$^{+2.4}_{-2.8}$~pc \citep{Galli21}. It is a  brown dwarf  with spectral type M7.5 and mass 0.04-0.05~M$_\sun$ \citep{Comeron00} which is still accreting at a low rate  \cite[$\dot{M}_{acc}$=10$^{-11.69}$~M$_\odot$year$^{-1}$;][]{Manara16}.
It is the first brown dwarf (BD) detected in X-rays \citep{Neuhauser98}. Its disk has been modeled in the past using Infrared Space Observatory (ISO) data up to 14.3 \mum requiring an optically thick, flared disk to account for the mid-IR flux \citep{Natta01}.Spitzer low resolution spectra show the detection of C$_2$H$_2$ in emission \citep{Pascucci09}.

Section~\ref{Observations} describes the observation and data reduction. In Section~\ref{DustFitting} and \ref{SlabModels} we describe the dust fitting and slab model retrievals used in this work. In Section~\ref{Results} we present our results on the dust content and molecular inventory. Finally we discuss our findings and present some conclusions in Sections~\ref{Discussion} and \ref{Conclusions}.

\section{Observations and data reduction}\label{Observations}

   \begin{figure}
   \centering
   \includegraphics[width=\hsize]{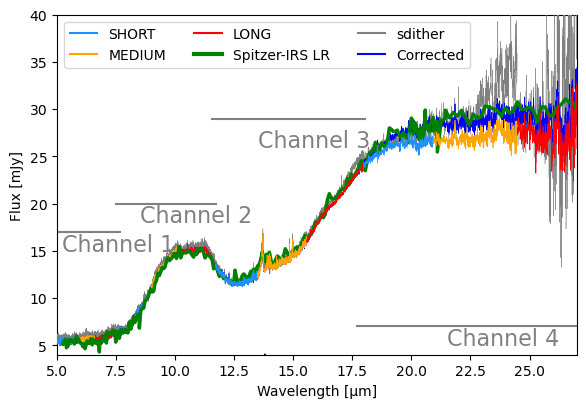}
      \caption{MIRI \chaha1 reduced spectra. Blue, yellow, and red colors denote the SHORT, MEDIUM and LONG bands respectively for each channel of the ddither approach.  The gray line shows the sdither reduction and the blue spectrum is the final flux-corrected one in Ch. 4. The green line corresponds to the Spitzer \chaha1 low resolution spectrum obtained from CASSIS \citep{Lebouteiller11}.}%
              \label{Fig:spec}
   \end{figure}

\chaha1 was observed with the MIRI on board the JWST on 15 August 2022 as part of the GTO program MINDS (ID: 1282, PI: Thomas Henning). The Medium Resolution Spectroscopy \cite[MRS;][]{Wells15,Argyriou23} mode was used, which involves four Integral Field Units (IFUs): channel 1 (4.9$-$7.65~$\mu$m), channel 2 (7.51$-$11.71~$\mu$m), channel 3 (11.55$-$18.02~$\mu$m), and channel 4 (17.71$-$27.9~$\mu$m). Each channel is composed of three sub-bands: SHORT (A), MEDIUM (B), and LONG (C) leading to a total of twelve wavelength bands (see Figure~\ref{Fig:spec}). Observations were performed with the target acquisition option using the same target and in FASTR1 readout mode with a point source 4-point dither pattern in the positive direction.  The total exposure time per sub-band
was 1232.12~s.

We implemented a hybrid pipeline for the reduction of our MIRI/MRS data. MINDS pipeline is publicly available\footnote{https://github.com/VChristiaens/MINDS} and for this reduction we used version 1.0.3. The pipeline relies on the
standard JWST pipeline \citep[v1.14.1;][]{Bushouse23} using CRDS context jwst\_1254.pmap, and is
complemented with routines from the Vortex Image Processing (VIP) package \citep{GomezGonzalez17,Christiaens23} to derive optimal-quality output spectra through background subtraction, bad pixel correction and centroiding options not available in the standard pipeline. The pipeline is structured around three main stages that are the same as in the JWST pipeline, namely Detector1, Spec2 and Spec3.
Stray light is corrected after the first stage (Detector1). From there, we considered two different approaches for background subtraction. In one case, to remove the background, we carried out a direct pairwise dither subtraction (hereafter referred to as ddither). This method is more suited to fainter sources where the resulting PSF overlap is minimal and reduces the noise level producing a higher signal-to-noise ratio (S/N) spectrum. However, the flux is lower than expected at the longest wavelengths due to some self-subtraction between minimally overlapping PSFs. In the second approach (hereafter referred to as sdither) we estimate and remove the background leveraging the four-point dither pattern to obtain a first guess on the background map, then we refine it using a median-filter which both smooths the background estimate and removes residual star signals from it. The result of this method is noisier and, as can be seen in Figure~\ref{Fig:spec}, there are some jumps. This is due to some structured bright background noise in the spectral frames of the long-ward end of the bands of channel~4 (and even in the LONG sub-band of channel~3). In the spectral frames of the beginning of each band the background seems to be well subtracted. Thus, we will assume that the beginning of each band of channel~4 in the sdither approach marks the absolute flux. We keep our higher S/N ddither reduction corrected in channel~4 to match the flux level at the beginning of each sdither band and linking it to the next band. In any case, the correction is only done for channel~4 and does not affect any molecular discussion since we have not found any molecule at wavelengths longer than $\sim$17~$\mu$m.
 
Spec2 was then used with default parameters, but the stray light correction and background subtraction were skipped. The outlier detection step in Spec3 is skipped, and replaced by a custom VIP-based bad pixel correction routine applied before Spec3. The bad pixels are identified through sigma-filtering, and corrected with a Gaussian kernel. VIP-based routines are then also used for the identification of the star centroid in the spectral cubes produced at stage 3. The centroid location is identified with a 2D Gaussian fit in a weighted mean image for each band's spectral cube, where the weights are set to be proportional to a proxy of the S/N of the source in each channel.

 The identified centroid locations are subsequently used for aperture photometry: the spectrum is extracted by summing the signal in a $1.5$ full width at half maximum (FWHM) aperture centered on the source, 2 FWHM for the ddither reduction, where the FWHM is equal to 1.22 $\lambda / D$, with $\lambda$ the wavelength and $D\sim6.5$m the diameter of the telescope.
Aperture correction factors are applied to account for the flux loss as presented in \citet{Argyriou23}.
 
The reduced spectra, using both background subtraction methods, are presented in Fig.~\ref{Fig:spec}. The different bands of the 4 channels in the ddither approach are highlighted in colors. We take the continuum-level from the  sdither approach, which is more accurate, and the higher S/N from the  ddither approach. To do so, we use a smooth polynomial to place the high S/N spectrum at the correct absolute flux level in a per band basis. The result is the blue line labeled as "corrected" in Fig.~\ref{Fig:spec}. Due to the faintness of the target,  the S/N is  poorer  in Ch. 4 and we decide to cut the spectrum at 26~$\mu$m. The spectral resolving power across the 5-25~$\mu$m spans from $R\sim$3500 at shorter wavelengths to $R\sim$1500 at longer wavelengths. The low-resolution $R\sim$60-127) Spitzer \chaha1 spectrum downloaded from  the Combined Atlas of Sources with Spitzer IRS Spectra  \citep[CASSIS;][]{Lebouteiller11} is also shown for comparison. The flux agreement of both sets of data is excellent.

\section{Dust fitting}\label{DustFitting}

The mid-IR excess emission observed in a young object arises from the hotter inner disk surrounding it. The shape of the continuum and the optically thin dust emission features generated by the dust grains in the disk surface layers can be used to probe the size and composition of the dust in the disk and therefore also the grain growth and crystallization \citep[e.g.,][]{Bouwman01}. That is, the degree of dust processing in protoplanetary disks can be estimated using silicate emission features present in the MIRI spectrum.

To fit the dust emission, we used  the Dust Continuum Kit \citep[DuCK;][]{Kaeufer24}, a dust fitting tool based on a two-layer disk model, as described by \citet{Juhasz09}. This allows us to calculate flux contributions from various components, including the central star, inner rim, optically thin disk surface, and optically thick mid plane. In the fitting process, we use absorption efficiencies (Q curves) of different dust species obtained with the Gaussian random field method \citep[GRF;][]{Min07} which generates individual grain shapes based on a Gaussian random field and calculates the average opacity over a range of individual particle shapes. We used a library of GRF Q curves including amorphous silicates MgOlivine \citep[Mg$_2$SiO$_4$,][]{Henning96} , MgPyroxene \citep[MgSiO$_3$,][]{Dorschner95}, and SiO$_2$ \citep{Spitzer60} and crystalline silicates forsterite \citep{Servoin73} and enstatite \citep{Jaeger98}.
 We used grain sizes of 0.1, 1, 2, 3, 4, and 5~$\mu$m. The fitting process follows \cite{Kaeufer24} and is described in detail in \cite{Jang24} with the only difference being that in the case of \chaha1 we allow for a temperature gradient in the disk surface instead of using a single temperature.

\begin{figure}
   \centering
   \includegraphics[width=\hsize]{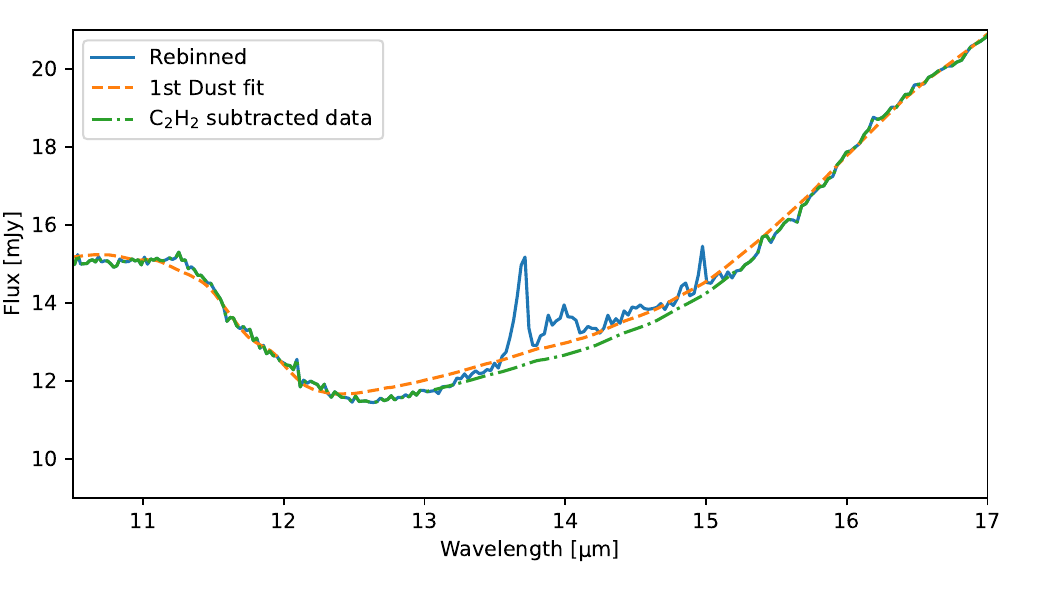}
      \caption{ Comparison between spectra before and after the residual gas subtraction. The blue line is the rebinned ($R \sim$ 500) MINDS data, and the orange line is the first dust fitting. The green line corresponds to the C$_2$H$_2$ subtracted data.  }
         \label{Fig:dustfitting}
   \end{figure}

\begin{figure}
   \centering
   \includegraphics[width=\hsize]{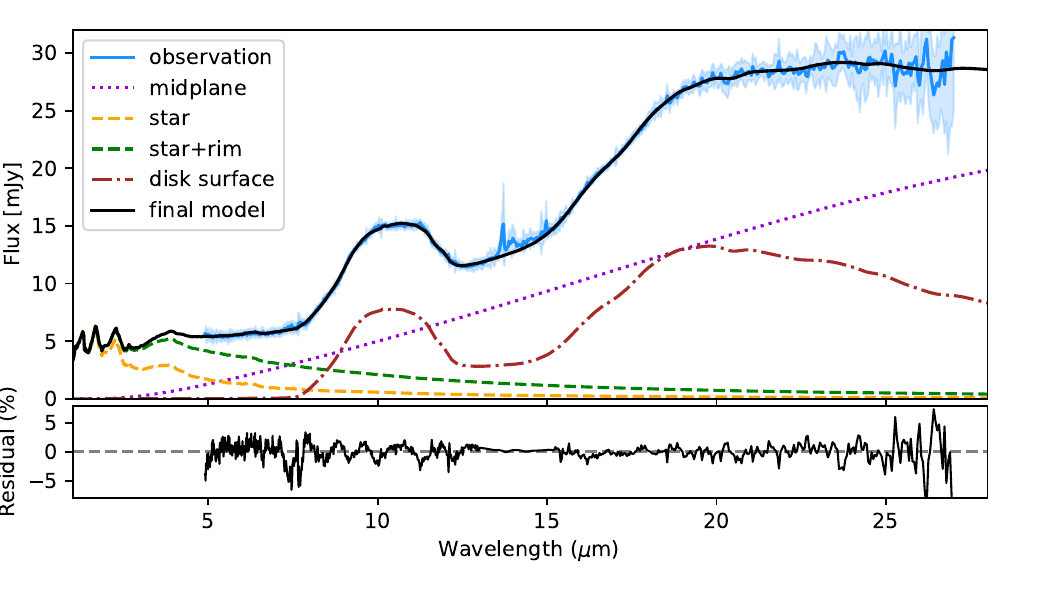}
      \caption{Dust fitting results after subtracting gas emission around 14~$\mu$m. The shaded area represents 1$\sigma$ uncertainties. The bottom panel shows the residuals of the fit. The effect of subtracting the C$_2$H$_2$ molecular emission can be seen at $\sim$14~$\mu$m.}
         \label{Fig:dust}
   \end{figure}

To fit the observed spectrum, we rebinned the MIRI data to have a constant spectral resolving power ($R \sim$ 500) to avoid the narrow gas emission lines (blue line in Figures~\ref{Fig:dustfitting} and ~\ref{Fig:dust}). Data beyond 26~$\mu$m is neglected due to significant noise. Given the gas-rich spectrum, particularly in the 14~$\mu$m wavelength range, we took additional steps to try to suppress the impact of gas emission on dust fitting.  We first fitted the spectrum, including the gas emission, and it resulted in a residual of approximately 2.5~mJy at the peak of the gas emission (orange dashed line in Figure~\ref{Fig:dustfitting}). Then the model is shifted down 0.3~mJy to make the data and model agree at 13~$\mu$m (green dot-dashed line in Figure~\ref{Fig:dustfitting}). Finally, the observed flux around 14~$\mu$m was subtracted from the shifted model. 
Spectra before and after the subtraction are compared in Fig.~\ref{Fig:dustfitting}, and the final fit is shown in Fig.~\ref{Fig:dust}. 

\section{Slab model fitting}\label{SlabModels}

In order to provide an inventory of molecules in the main planet-forming region of the disk of \chaha1, that is inside a few au,  we use  zero-dimensional (0D) slab models calculated with the {\sc ProDiMo} code \citep{Woitke09, Kamp17, Woitke16}. The 0D slab model fitting  is a simplified way to describe a layer (or slab) of material with uniform properties, that is, it assumes a homogeneous column of gas that radiates with a fixed temperature and a constant column density.  A grid of  slab models, one for each of the molecules investigated, was generated in  \cite{Arabhavi24} following  previously described methods \citep{Tabone23, Grant23}. We use the synthetic spectra produced in \citet{Arabhavi24} where line positions, Einstein A coefficients, statistical weights, and partition functions, were taken from the HITRAN 2020 database  \citep{Gordon22}, the GEISA database  \citep{Delahaye21} and \cite{Arabhavi24} for C$_3$H$_4$ and C$_6$H$_6$. An overview of the molecular data used for calculating the mid-IR spectra is shown in Table 2 of \cite{Henning24}. Note that we use the LTE treatment for the molecules presented in that table. These synthetic spectra are available for a wide range of ambient conditions (e.g. gas temperature, gas column density). In particular, we  sample the temperature in steps of 25 K from 25 K to 1500 K and column densities in steps of $1/6$~dex from 10$^{14}$~ cm$^{-2}$ to 10$^{24.5}$~cm$^{-2}$. For C$_6$H$_6$ and C$_3$H$_4$  the temperature is limited to $\leq$600 K because of incompleteness of the laboratory data \citep{Arabhavi24}.

The slab models outlined above provide total line fluxes in a certain wavelength range per molecule and account for the distance to \chaha1 \citep[191.8~pc,][]{Galli21}. For comparison with the observed data the MIRI spectrum needs to be continuum subtracted. The continuum and continuum subtracted spectrum are shown in Figures~\ref{Fig:App:cont} and ~\ref{Fig:App:conttest}. See Appendix~\ref{App:Continuum} for a detailed description on the continuum definition. We then convolve the slab models with the appropriate spectral resolution of the instrument, in this case R$\sim$2500-3500 depending on the channel \citep{Labiano21, Pontoppidan24}. Finally, the convolved model spectrum is resampled to have the same wavelength grid as the observed spectrum.

We employ $\chi^2$  minimization with the prodimopy (v2.1.4)\footnote{Available at https://gitlab.astro.rug.nl/prodimo/prodimopy}  tool to identify the best fitting slab model. Our investigation encompasses the parameter space of column density ($N$), temperature ($T$), and the equivalent radius of the emitting area ($R$). The emitting area radius acts as a scaling factor for achieving the absolute flux level while the column density and temperature capture the relative flux levels of lines or the spectral shape. 
We define very restricted wavelength windows for the $\chi^2$-minimization trying to avoid the contribution from other molecules while still containing features that help to constrain the fit (e.g., optically thin and temperature-sensitive lines). These windows are shown in Figure~\ref{Fig:models}. The fitting is done in an iterative way, one molecule at a time, to further reduce contamination from other species going from the strongest line emission contributor to the weakest, subtracting the fit from the spectrum for each molecule before starting the fitting of the next molecule. However we note that some molecules present quite some overlap in their model spectrum which we do not take into account in our fitting procedure. This procedure results in some differences to a simultaneous fit of all molecules but when no pseudo-continua exists there is strong overlap \citep{Kaeufer24}. In cases where the emission overlap is complete (C$_3$H$_4$ and C$_4$H$_2$, and C$_2$H$_2$ - see Sec.~\ref{Results}) we have fitted the two molecules emitting in the same area simultaneously. In these cases the free parameters are N$_1\times$T$_1\times$R$_1\times$N$_2\times$T$_2\times$R$_2$.

\begin{figure*}
   \centering
   \includegraphics[width=\textwidth]{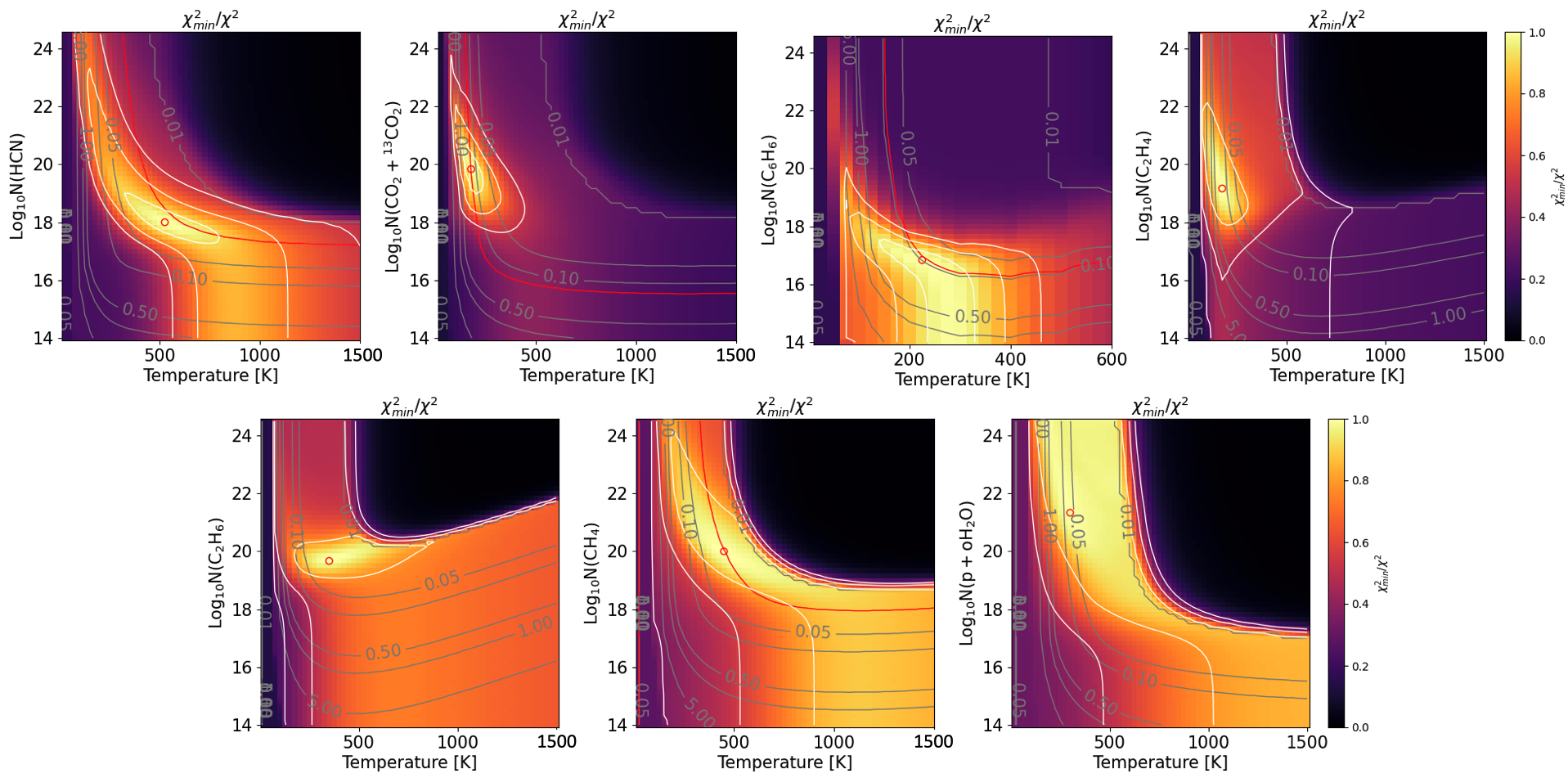}
      \caption{ $\chi^2$ maps for HCN, CO$_2$ +  $^{13}$CO$_2$, C$_6$H$_6$, C$_2$H$_4$, C$_2$H$_6$, CH$_4$, and  water. The color scale shows the $\chi^2_{min}$/$\chi^2$. The best fit model ($\chi^2_{min}$/$\chi^2$ = 1) is marked with a red circle. A red contour denotes the emitting region radius as listed in Table~\ref{Tab:molecules}. Gray contours show the emitting radii in au while white contours represent the 1$\sigma$, 2$\sigma$, and 3$\sigma$ levels. 
}
         \label{Fig:xis}
   \end{figure*}

In that way, the order for the fitting has been C$_2$H$_2$ - HCN - CO$_2$ - C$_6$H$_6$ - C$_3$H$_4$ - C$_4$H$_2$ - C$_2$H$_4$  - C$_2$H$_6$ - CH$_4$ - H$_2$O. 
The $\chi^2$ maps are presented in Figure~\ref{Fig:xis} and show the 1$\sigma$-, 2$\sigma$-, and 3$\sigma$- contours. However, these are not a good approximation of the error because of the stepwise procedure used and the molecular emission overlap. In Table~\ref{Tab:errors} we show the estimated rms noise level, $\sigma$ of each band where molecular emission was found. To estimate this uncertainty we use the JWST Exposure Time Calculator (ETC\footnote{https://jwst.etc.stsci.edu}) to obtain an estimate of S/N as described in \citet{Jang25}.
Another way to measure the errors of the derived parameters is to repeat the fitting in a different order. In this way we find changes in the derived temperatures and column densities of 75~K and 0.5 orders of magnitude respectively. Finally we used the following criteria to determine if a molecule detection is robust: if at least two features are detected, if the main feature is stronger than the 3$\sigma$-uncertainty, and if the same molecule is detected if we change the order of the fitting we consider it a firm detection. If any of those criteria fail we will declare the detection tentative. We chose to require at least two features for a robust detection to adopt a conservative approach, and we did not impose a three-feature requirement because for some molecules, such as for C$_6$H$_6$, the third feature may be challenging to detect within the explored range of temperature and column density. Note that even in the case of a firm detection, there are cases where the parameters cannot be well constrained because the fit is degenerate (see Fig.~\ref{Fig:xis}). In these cases we do not provide a value for the parameters and instead will just list the molecule as "Detected".

We performed the exercise of determining different continuum levels and repeat the molecular fits for each of them in order to evaluate the differences. The same molecules were detected in all cases. The results of this test can be seen in Appendix~\ref{App:fits}.

\begin{table*}[h!]
\caption{Median continuum flux, estimated S/N from the ETC,  and 1$\sigma$- noise for each band where molecules are detected.}             \label{Tab:errors}    
\centering               
\begin{tabular}{l c c c c}
\hline\hline                

Band & Median flux & ETC S/N & 1$\sigma$- & Molecules \\

[$\mu$m]         &   [mJy]	&	&   [mJy]    &		\\ 
\hline                     
1C (6.63 - 7.65) &	5.88	& 90	&	0.07	&	H$_2$O, CH$_4$ \\
2C (10.13 - 11.70) &	15.00 & 200		&	0.08	&	C$_2$H$_4$ \\	
3A (11.70 - 13.47) &	11.81 & 170		&	0.07	&	C$_2$H$_6$ \\	
3B (13.47 - 15.57) &	13.92 & 170		&	0.08	&	C$_2$H$_2$, HCN, \\
			     &		&	&		&	C$_6$H$_6$, CO$_2$, \\

3C (15.57 - 17.98) &	20.17 & 220		&	0.09	&	C$_3$H$_4$, C$_4$H$_2$, \\
			     &		&	&		&	CH$_3$ \\

\hline                      
\end{tabular}
\end{table*}

\section{Results}\label{Results}

The first thing to note is the good agreement of the JWST spectrum with the Spitzer data (Fig.~{\ref{Fig:spec}}), taken in 2005, showing the absence of detectable variability in this wavelength range. This is in contrast to the finding of aperiodic variability with amplitude of $\sim$0.4~mag at optical wavelengths which is attributed to accretion and disk properties \citep{Cody14}.

The much higher spectral resolution of MIRI-MRS compared to previous Spitzer data reveals a plethora of molecular features (Fig.~\ref{Fig:models}). C$_2$H$_2$ emission at 13.7~$\mu$m is particularly strong, consistent with
earlier findings that this molecules emission is enhanced in disks around brown dwarfs and VLMS \citep{Pascucci09}.  The other molecular features can be ascribed to several hydrocarbon molecules (C$_6$H$_6$, C$_3$H$_4$, C$_4$H$_2$, C$_2$H$_4$, C$_2$H$_6$, CH$_4$, and CH$_3$) plus CO$_2$,  HCN, H$_2$, and H$_2$O.  In addition, one $^{13}$C-substituted isotopologue each for acetylene and carbon dioxide (hereafter $^{13}$CCH$_2$ and $^{13}$CO$_2$) have been found. The multitude of detected carbon bearing molecules points to a carbon-rich inner disk environment as has been seen in TWA 27 A \citep{Patapis25} and in some VLMS \citep{Tabone23, Arabhavi24, Kanwar24}. However this is the only case in which water is clearly detected on top of all the hydrocarbons.

\subsection{Dust content}\label{Results:Dust}

The dust grains typical to  the interstellar medium (ISM) are sub-micron ($\sim$0.1~$\mu$m) sized, amorphous silicate grains \citep[e.g.,][]{Kemper04,Bouwman01,Apai04} which  show  a  single-peaked, relatively  sharp  and  strong, unstructured feature peaking at 9.8~$\mu$m. In contrast, what we see  in the \chaha1 spectrum is a broader and flat topped emission feature typical of larger amorphous grains indicating that some dust processing has already happened within the very first million years of disk formation. 
Less prominent forsterite features can be seen around 16~$\mu$m. 
No PAH features are found in agreement with possible PAH destruction by EUV or X-ray photons, as discussed by \cite{Siebenmorgen10}.

In agreement to what can be extracted from looking at the spectrum, our dust fitting reveals that dust is dominated by large $\sim$4~$\mu$m  sized amorphous silicates (>50$\%$ mass fraction of MgSiO$_3$). This shows that grain growth is already underway at the age of  $\sim$1.5~Myr as has been seen in previous works \citep{Kessler07,Furlan09, Ribas17, Grant18}. We also find a lower mass fraction of smaller grains of MgOlivine. Notably, amorphous SiO$_2$ is detected only marginally. For crystalline silicates, we only see large grains of forsterite (13$\%$ of $\sim$5~$\mu$m size) which have very weak and flat features. No enstatite was detected in contrast to what is seen in other BD disks \citep{Apai05,Merin07}.

\subsection{Molecular inventory}\label{Results:Gas}
\begin{figure*}
   \centering
   \includegraphics[width=\hsize]{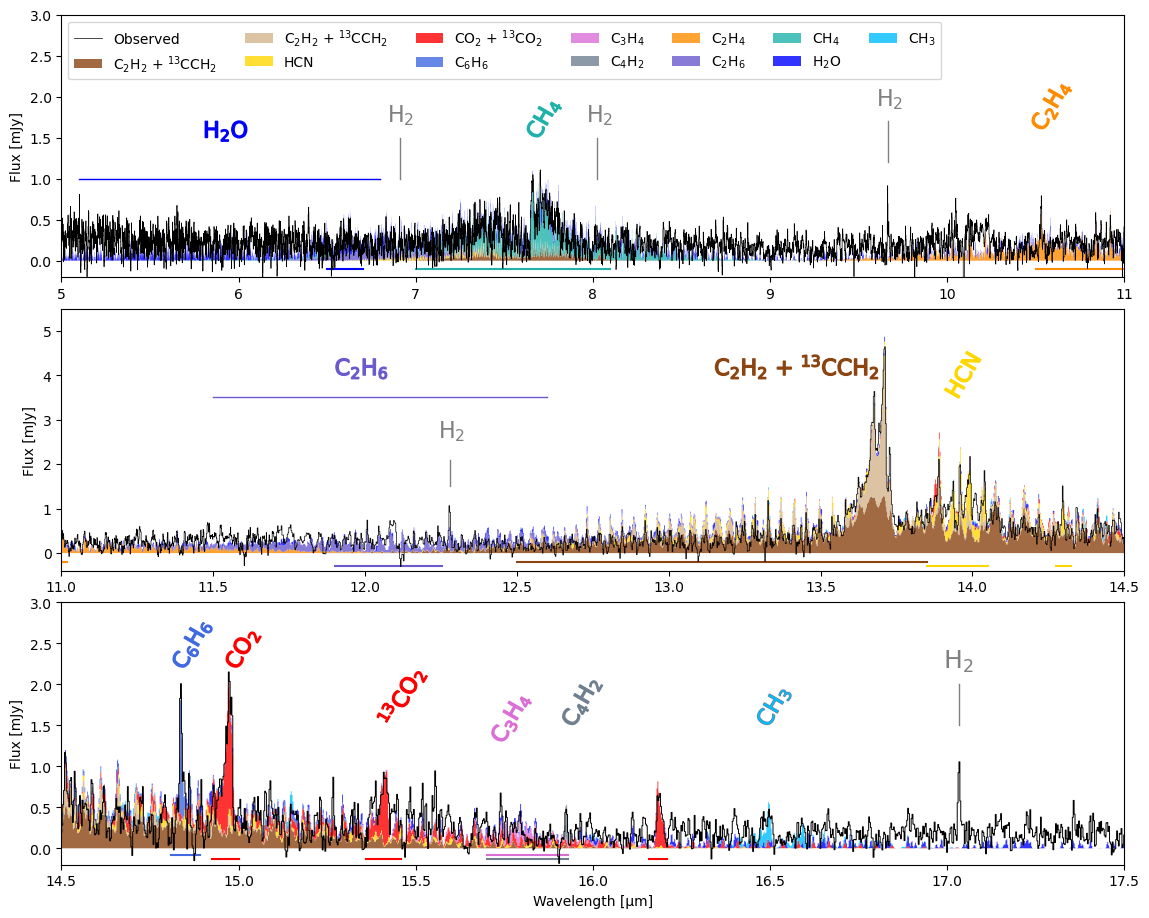}
      \caption{\chaha1 spectrum, with the JWST-MIRI continuum subtracted data (black) compared to the stacked emission from the slab models for C$_2$H$_2$ + $^{13}$CCH$_2$, HCN, CO$_2$ + $^{13}$CO$_2$, C$_6$H$_6$, C$_3$H$_4$, C$_4$H$_2$, C$_2$H$_4$, C$_2$H$_6$, CH$_4$, H$_2$O, and CH$_3$. The parameters of the slab models shown can be found in Table~\ref{Tab:molecules} and ~\ref{Tab:App:molecules}. The H$_2$ lines are also marked. Horizontal lines represent the windows in which the $\chi^2$ for each fitting has been evaluated. Note that the y axis are different for each panel.}
         \label{Fig:models}
   \end{figure*}

\begin{figure}
   \centering
   \includegraphics[width=\hsize]{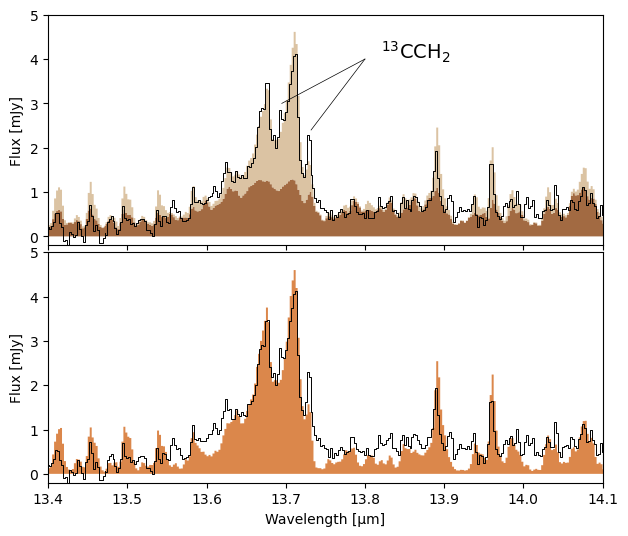}
      \caption{Zoom into the 13.7~$\mu$m region. The upper panel shows the fit using two C$_2$H$_2$ slab models with optically thick (dark brown) and thin (light brown) emission. Two peaks at 13.695~$\mu$m and 13.732~$\mu$m associated with $^{13}$CCH$_2$ can be clearly distinguished. The lower panel shows the best fit when using just one slab model (T = 325~K, N=3.1$\times$10$^{18}$~cm$^{-2}$, and R=0.07~au). }
         \label{Fig:c2h2}
   \end{figure}

\begin{figure}
   \centering
   \includegraphics[width=\hsize]{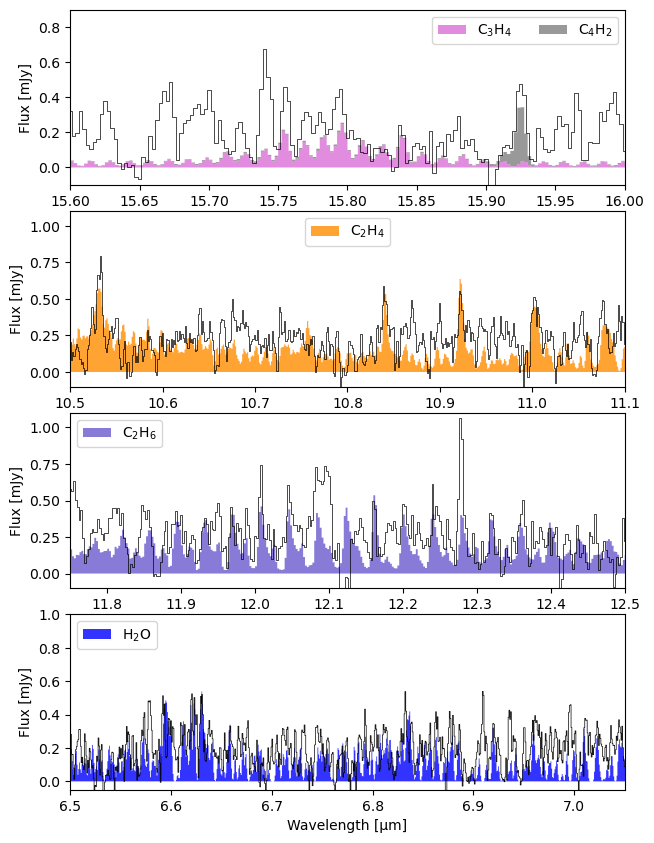}
      \caption{Blows up for slab model fits of C$_3$H$_4$ and C$_4$H$_2$, C$_2$H$_4$,  C$_2$H$_6$,  and H$_2$O. Note that the slab models in the first panel are stacked one on top of the other. The parameters of the slab models shown can be found in Table~\ref{Tab:App:molecules}}
         \label{Fig:fits}
   \end{figure}

\begin{figure}
   \centering
   \includegraphics[width=\hsize]{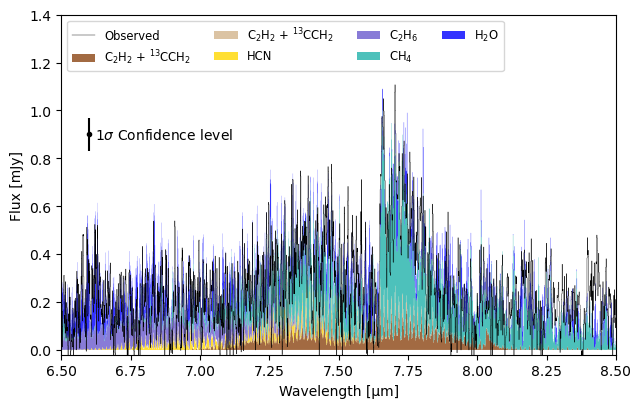}
      \caption{Zoom into the 7.5~$\mu$m region. Stacked slab model fits of HCN, CH$_4$,  C$_2$H$_2$, C$_2$H$_6$,   and H$_2$O are shown. The dark and light brown colors for C$_2$H$_2$ + $^{13}$CCH$_2$ represent the optically thick and thin models respectively.} 
         \label{Fig:ch4}
   \end{figure}

The best fit models for each molecule are presented in Figure~\ref{Fig:models} together with the windows used for the $\chi^2$ minimization. Blows up for some molecules can be seen in Figures ~\ref{Fig:c2h2} -~\ref{Fig:ch4}. The best-fitting parameters are summarized in Table~\ref{Tab:molecules}.

\begin{table}[h!]
\caption{Temperatures, column densities and emitting radii of the detected molecules obtained
from the slab model fits.}      
\label{Tab:molecules}      
\centering                        
\begin{tabular}{l c c c}     
\hline\hline                
Species & $T[K]$ & N[10$^{18}$cm$^{-2}$] & R[au] \\   
\hline                        
C$_2$H$_2$ + $^{13}$CCH$_2$ & 300 & 316 & 0.06 \\    

C$_2$H$_2$ + $^{13}$CCH$_2$ & 300 & 0.7 & 0.07  \\  

HCN & 525 & 1   & 0.02  \\

CO$_2$ +  $^{13}$CO$_2$ & 175 & 68     & 0.15  \\

C$_6$H$_6$ & 225 &0.07\tablefootmark{b}    & 0.09   \\

C$_3$H$_4$ & \multicolumn{3}{c}{Tentative Detection}  \\ 

C$_4$H$_2$ &  \multicolumn{3}{c}{Detected}    \\

C$_2$H$_4$ &  \multicolumn{3}{c}{Detected}  \\

C$_2$H$_6$ &  \multicolumn{3}{c}{Tentative Detection}  \\

CH$_4$ & 450 & 100 & 0.02  \\      

H$_2$O   &  \multicolumn{3}{c}{Detected}   \\

CH$_3$ &  \multicolumn{3}{c}{Detected}     \\

\hline                              
\end{tabular}
\tablefoot{Typical uncertainties are of order  75~K and 0.5 orders of magnitude for temperature and column density, respectively.\\
\tablefoottext{b}{Parameter not well constrained.} 
}
\end{table}

{\bf C$_2$H$_2$ and $^{13}$CCH$_2$:}
As mentioned before, the Q branch of C$_2$H$_2$ at 13.7~$\mu$m associated with the $\nu_5$ bending mode is the most prominent molecular feature in the entire MIRI-MRS spectrum. This is also the only feature that has been seen in previous low-resolution data from Spitzer. At a spectral resolving power of $\sim$~2500, MIRI-MRS  also reveals the shortward peaks at 13.63 and 13.68~$\mu$m due to hot bands that were blended with the main peak in lower resolution Spitzer spectra. 
Previous observations of VLMS have shown that, in some cases, an optically thick component forms a pseudo-continuum of molecular emission  \citep[e.g.,][]{Tabone23, Arabhavi24, Kanwar24}. Thus we treated the C$_2$H$_2$ molecule as if it was two different species fitted simultaneously but we limit the parameter space of the column density of the first model to larger column densities (N>10$^{19}$~cm$^{-2}$, which is when the molecular pseudo-continuum appears for the range of temperatures explored) and the second model to lower column densities.
We assume that the temperature is the same for both models and let it vary between 250 and 450~K.
Figure~\ref{Fig:c2h2}  shows that the overall shape of the 13.7~$\mu$m feature can be well fitted with a model with a larger column density which reproduces well the molecular pseudo-continuum and another model with lower column density which reproduces the peaks of
the $Q$-branches. The lower panel Fig.~\ref{Fig:c2h2} shows, for comparison, the best fit model when using just one slab for the fit. In addition, two peaks at 13.695 and 13.732~$\mu$m associated with $^{13}$CCH$_2$ can be clearly seen.  Our fit does include the contribution of $^{13}$CCH$_2$  assuming a C$_2$H$_2$/$^{13}$CCH$_2$ ratio of 35 \citep{Woods09}  but the models cannot fully reproduce the observed spectrum due to incomplete molecular data. Proper modeling must await more complete molecular spectroscopy for this C$_2$H$_2$ isotopologue.

{\bf HCN:}
The ro-vibrational band from the fundamental $\nu_2$ bending mode of HCN at 14.0~$\mu$m and the hot band at 14.3~$\mu$m are both easily distinguished  in Fig.~\ref{Fig:models} and are well fit by a slab of gas at a warmer temperature (in the range 300-800~K).  However we note that  there are strong P-branch emission lines of C$_2$H$_2$ in the same wavelength range as HCN and C$_2$H$_2$ has been fitted first  and the result subtracted from the MIRI spectrum. We have done the exercise of fitting first HCN and the obtained results differ by ~50 -  100~K, a smaller error than  the degeneracy that can be seen  in the $\chi^2$ map in Fig.~\ref{Fig:xis}. 

\citet{Pascucci09}  found that the the emission ratio of HCN relative to C$_2$H$_2$ in the disk surface of cool stars was weaker compared to the T Tauri disks. They could not detect HCN in the Spitzer spectrum of \chaha1 but, with the MIRI spectrum, we  can confirm that the C$_2$H$_ 2$ /HCN line flux ratio is larger than unity as expected for  cool stars instead of smaller than unity as seen in T Tauri stars. The HCN emission in \chaha1 is much larger, relative to C$_2$H$_2$ than in the VLMS 2MASS J16053215-1933159 \citep[hereafter J1605;][]{Tabone23} or ISO-ChaI~147 \citep{Arabhavi24} that have a stronger molecular continuum and higher stellar luminosities.

{\bf CO$_2$ and $^{13}$CO$_2$:}
Figure~\ref{Fig:models} includes the fit to the CO$_2$ bending mode at 14.98~$\mu$m that is clearly detected.  The Q branch of $^{13}$CO$_2$ at 15.4~$\mu$m is also clearly detected. The isotopologue $^{13}$CO$_2$ has been also recently detected in other protoplanetary disks \citep[e.g.,][]{Grant23, Arabhavi24, Kanwar24, Vlasblom24}. We use a slab model with both species included using the standard CO$_2$/$^{13}$CO$_2$ ratio of 70 from the interstellar medium \citep{Milam05}. The best fit indicates cooler gas at 175~K and larger radii ($\sim$0.15~au) than the other molecules. The $\chi^2$ map in Fig.~\ref{Fig:xis} shows that the gas temperature is well constrained below 250~K.

{\bf C$_6$H$_6$:}
This is the largest molecule that we have detected. We identified two features around 14.85~$\mu$m to be Q branches of the
fundamental and hot bending mode $\nu_4$ of benzene, C$_6$H$_6$, presented in Fig.~\ref{Fig:models}. Their relative intensity is sensitive to temperature and indicates gas at $T$=225~K. However, we note that in this case, even when C$_6$H$_6$ is clearly detected, the column density  is degenerate (see Fig.~\ref{Fig:xis}) and our best fit value shown in Table~\ref{Tab:molecules} is likely an upper limit.

{\bf C$_3$H$_4$ and C$_4$H$_2$:} The CH bending mode $\nu_9$ at $\sim$15.80~$\mu$m for  C$_3$H$_4$ is tentatively detected (Fig.~\ref{Fig:fits}) while the CH bending mode $\nu_5$ at 15.92~$\mu$m for C$_4$H$_2$ stands out more clearly. Since the emission of both species almost completely overlap, we fitted both molecules simultaneously leaving the column density as a free parameter and letting the temperature vary between 175 and 350~K. In Fig.~\ref{Fig:fits}, we show the MIRI spectrum in the range of wavelengths where C$_3$H$_4$ and C$_4$H$_2$ emit. In the case of C$_3$H$_4$, the strength of the features to fit are similar to the residuals and,  according to our criteria, we consider it a tentative detection. For C$_4$H$_2$ there are several features and the strength of the main one is over 3$\sigma$; so we consider it a real detection.

{\bf C$_2$H$_4$:} A zoom into the 10-11~$\mu$m region can be seen in Fig.~\ref{Fig:fits} showing the feature at 10.53~$\mu$m. Note that while the $\chi^2$ map in Fig.~\ref{Fig:xis} is quite well constrained, this spectral region can be affected by the broad silicate emission and thus the uncertainty in the derived parameters can be higher that that obtained from the $\chi^2$ maps. In fact, the slab model fits well the  main peaks of C$_2$H$_4$ but following our conservative approach, and because there are multiple unidentified wiggles between the five peaks of C$_2$H$_4$ that are almost as bright than few C$_2$H$_4$ features, we claim a tentative detection.

{\bf C$_2$H$_6$:} The emission of C$_2$H$_6$ can be seen in Fig.~\ref{Fig:fits} where the line at 12.279~$\mu$m is H$_2$~S(2). We set a window to evaluate the fit so it avoids this line. However, since the emission of this molecule is of order of the residuals and it is affected by the emission of C$_2$H$_2$ and  C$_2$H$_4$, we consider it a tentative detection.

{\bf CH$_4$:}
Figure~\ref{Fig:models} shows the Q branch of the $\nu_4$ mode of CH$_4$ around 7.66~$\mu$m and Figure~\ref{Fig:ch4} is a zoom into that region.  CH$_4$ was previously seen in the GV Tau N (0.8 \Msun) T Tauri disk in absorption \citep{Gibb13} and more recently in emission in the very low mass disks of J1605 \citep[tentative detection:][]{Tabone23},  ISO-ChaI 147 \citep{Arabhavi24}, Sz28  \citep{Kanwar24}, and TWA 27 A \citep{Patapis25}. In the case of ISO-ChaI 147, CH$_4$ is the molecule with the largest column density and it forms a molecular continuum. We note that C$_2$H$_2$ and HCN also have some emission lines in this region, but cannot reproduce by themselves this broad feature. 

{\bf H$_2$O:} Figure~\ref{Fig:fits} shows the detection of the ro-vibrational band of water in the 6~$\mu$m region. For this detection we have used the detection criteria used in \cite{Arabhavi25a}: a detection is real if there are at least 5 lines detected in the wavelength range between 6.5 and 6.7~$\mu$m. The fit is degenerate and, while we can provide a probable temperature of $T$=500~K, the lines in the region are not sensitive to column density. This is the first BD disk for which ro-vibrational water is detected and the only source for which several hydrocarbons have been detected along with water \citep{Arabhavi25a}.  
We do not detect the pure rotational water emission seen in some VLMS between 16 and 18~$\mu$m. 

{\bf CH$_3$:} Figure~\ref{Fig:models} shows our CH$_3$ detection. We used the molecular data from \cite{Helmich96}, however, it is incomplete and we could not fit the emission from this molecule in the same way as we did for the other species. We show a model with 400~K to confirm the detection.

   \begin{figure*}
   \centering
    \includegraphics[width=\hsize]{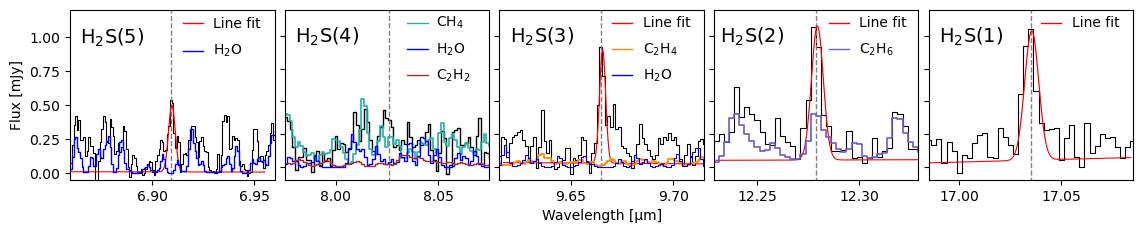}
      \caption{ Zoom into the  positions of the H$_2$ pure rotational lines in the  \chaha1 MIRI spectrum. The position of the H$_2$ lines are marked with dashed lines.The red profiles show line fitting using a Gaussian function. The emission from other molecules is marked in color in each panel and its is based on slab models reported in the Tables~\ref{Tab:molecules} and ~\ref{Tab:App:molecules}.     }
         \label{Fig:h2}
   \end{figure*}

{\bf H$_2$:}
In addition to the many other features, several H$_2$ pure rotational ($\nu$ = 0 -- 0) lines are clearly identified in the MIRI-MRS spectrum (see Figs.~\ref{Fig:models} and \ref{Fig:h2}),  in particular, S(1)-S(3) and S(5) are clearly detected,  while S(4) is only tentative. The emitting region seems to be unresolved, however some extended emission is barely noticeable in Figure~\ref{Fig:App:h2extended}. Hence, the emission is most probably coming from the very inner, warm region of the disk. 

As can be seen in Figure~\ref{Fig:h2} S(4) and S(5) are heavily contaminated by the emission of other molecules, mainly CH$_4$ and water, respectively. S(3) also falls on top of the C$_2$H$_4$ emission. Note that while we plot a slab model of the C$_2$H$_4$ emission in Figure~\ref{Fig:h2}, mid panel, the fit of the model was not well constrained. The same is applicable to the S(2) transition (that falls on top of the C$_2$H$_6$ emission). An in depth modeling of these emissions is needed in order to separate the emission of H$_2$ and calculate accurate fluxes. 

To measure preliminary line fluxes we follow \cite{Franceschi24} who performed a similar analysis for J1605. The H$_2$ lines were fitted in the continuum subtracted spectrum using a Gaussian profile. This was done using the Levenberg-Marquardt least-squares minimization. The H$_2$ line flux was then calculated by integrating the Gaussian component.
The line profiles and the best-fit profiles are shown in Fig.~\ref{Fig:h2}, and the line parameters are reported in Table~\ref{Tab:H2fluxes}.

\begin{table}[h!]
\caption{Measured parameters of the identified pure rotational molecular hydrogen lines.}    
\label{Tab:H2fluxes}     
\small
\centering   
                      
\begin{tabular}{l c c c}    
\hline\hline              
Transition & Line center& FWHM &Integrated flux\\   
                 & [$\mu$m] &  [$\mu$m]& 10$^{-16}$ [erg s$^{-1}$ cm$^{-2}$]\\
\hline                       
S(1) &  17.035 & 0.008& 0.9$\pm$0.1 \\    

S(2) &  12.280 & 0.007& 1.5\tablefootmark{a}  \\  

S(3) &  9.665   & 0.004& 1.1$\pm$0.1  \\

S(5) & 6.910    & 0.004& 1.3\tablefootmark{a}   \\
\hline                              
\end{tabular}
\tablefoot{
\tablefoottext{a}{The line is on top of considerable emission coming from other molecule and hence the flux can be considered an upper limit.} 
}
\end{table}

Finally, there are clear emission features in Fig.~\ref{Fig:models} at $\sim$10\mum and long ward of the CH$_3$ ($\sim$16.5\mum) emission, which correspond to emission that we have not been able to ascribe to any molecule.

\section{Discussion}\label{Discussion}

\subsection{Dust in the disk of \chaha1}

The features present in the MIRI spectrum probe dust emission originating essentially from large silicate dust grains located in the optically thin surface layers of the inner disk indicating that grain growth is already underway at an age of  $\sim$1.5 Myr. The dominant grain size is found to be $\sim$4 \mum which is in the larger side if we compare it with the sample of T Tauri stars from several associations studied by \cite{Oliveira11}. If we compare \chaha1 with the other disks around BDs which have been observed with MIRI, that is TWA 27 A \citep{Patapis25} and J0438 \citep{Perotti25}, the first thing to note is the difference in the dust features. TWA 27 A  shows a complete absence of silicate features implying that either it is dust-depleted or the grains have grown to large sizes ($>$5 \mum).  In either case the dust opacity will be very low and we can see deeper into the disk, closer to the mid plane.
J0438 has a highly inclined disk and thus some of the silicate features turn into absorption, resulting in a smaller 10\mum feature. Besides the combined (absorption and emission) silicate feature, the overall spectrum shape is smooth, indicating dominant amorphous silicates and no significant crystalline features. Trying to find a similar spectral shape to that of \chaha1 we will compare it also with Sz28 \citep{Kanwar24,Kaeufer24}. Sz28 is a VLMS (M5.5) with a MIRI spectrum and it does show  a 10 \mum feature though less prominent than that of \chaha1. Its dust emission is also dominated by large ($\sim$5 \mum) grains.  This aligns with the finding that the region of a protoplanetary disk traced by the 10 \mum silicate emission features is linked to the luminosity of the central object \citep{Kessler07}. Consequently, the grain sizes inferred from these features may primarily reflect the location of the silicate emission zone, rather than the overall properties of the disk’s silicate dust. In VLMs and BDs, this implies we are observing dust from a region closer to the star compared to their higher-mass counterparts.
The crystallinity fraction derived is in the range of that observed by \cite{Oliveira11}. Sz28 shows a larger fraction (20\% vs. 13\% for \chaha1) than \chaha1. In addition they find a different population of crystalline dust in the warm zone ($\le$1 AU) emitting at $\lambda\sim$10 \mum and the  colder region emitting at $\lambda> $16 \mum. We do not see any evidence, when fitting the dust, of an inhomogeneous composition of the dust. Further discussion on dust properties for a sample of VLMSs and BDs which include \chaha1 can be found in \citet{Jang25}. 

As a protoplanetary disk evolves, dust grains are believed to grow in size. As they become larger, they are more strongly influenced by gravity, causing them to settle toward the disk’s mid plane and deplete from the upper disk layers \citep{Dullemond04}. During this process, the strength of the silicate emission feature diminishes. In this scenario, \chaha1 may be slightly less evolved than Sz28, still with a flared disk in which dust settling into the mid plane is still in its initial steps. TWA 27 A would be in a more evolved stage when substantial grain growth and dust sedimentation have occurred in the disks.

\subsection{Volatile content}

The large number of carbon-bearing molecules (nine hydrocarbons plus HCN) detected in \chaha1 points to a carbon-rich inner disk environment. Nevertheless, we also find the presence of H$_2$O, CO$_2$, and $^{13}$CO$_2$. The presence of water lines at $\sim$7~$\mu$m points to a reservoir of hot water in the inner disk. 

Regarding the temperature differences, hydrocarbons in general  have temperatures in the 225-450 K range as have been seen  in other MINDS VLM objects and points to a common reservoir for all of them. They emit from a confined area within 0.1~au from the host star while CO$_2$ seems to be emitted from slightly further out or deeper in the disk. On the contrary, HCN is much warmer, probably closer or higher in the disk. This trend in the CO$_2$ is seen also in Sz28 and other VLMS \citep{Arabhavi24}. \citet{Tabone23} suggest that a gap in the disk might prevent O-rich icy grains from moving inward, resulting in a high C/O ratio in the inner disk, which enables strong mid-IR emission. If we accept this idea, the outer disk could still maintain a standard C/O ratio, potentially explaining the observed CO$_2$ emission and the fact that it is systematically cooler than the hydrocarbons (at least for these objects). Recently, \cite{Kanwar24a, Kanwar25} were able to reproduce the molecular emission from different species simultaneously with a thermo-chemical disk model with a gap. In that model, hydrocarbons come from inside the gap, while CO$_2$ emission comes from outside the gap.

If we compare \chaha1 with the other BD disks observed with MIRI, that is J0438 and TWA 27 A, J0438 is peculiar because it resembles the spectra of higher mass T Tauri disks with a volatile reservoir lacking hydrocarbons (except for acetylene, C$_2$H$_2$) and dominated by water. TWA 27 A is hydrocarbon-rich, all detected molecules contain carbon, while CO$_2$  is the only oxygen-bearing molecule. The diversity of hydrocarbons indicates that the C/O ratio is larger than unity. This is also the case of Sz28. In both cases C$_2$H$_2$ has a very high column density and forms a molecular pseudo continuum (as happens with \chaha1) with column densities higher than that of \chaha1.  The diversity of molecules in the three objects (TWA 27 A, Sz28, and \chaha1), is similar and all of them show a carbon rich inner disk, however, the dominant species differ. For example TWA 27 A shows C$_2$H$_4$ as the dominant hydrocarbon while in Sz28 this molecule is not detected and for  \chaha1 it is only a tentative detection. Among the four objects \chaha1 is the most diverse in molecular species. It presents a similar amount of hydrocarbon molecules detected and, in addition, it also has water. However, note that \chaha1 is the only source for which the spectrum at short wavelengths, $<$7 \mum, is not dominated by stellar photospheric absorption (that could mask the presence of warm water in the other sources). In addition, \citep{Arabhavi25a} show that bright
hydrocarbon emission can outshine emission from large column
densities of water.

\citet{Mah24} found that in VLMSs, very short viscous timescales and close-in ice lines contribute to a high gaseous C/O ratio after about 2 Myr. By including disk substructures, they observed that for low viscosities, the inner disk can remain water-rich for extended periods (at least 5 Myr), regardless of the gap’s depth or formation timing. This could explain water-rich sources like J0438. If, after 2 Myr, the disk starts shifting toward a high C/O ratio, objects like \chaha1—with faint water signatures, various hydrocarbons, and shallow molecular continuum—could be in the midst of this transition. Sz28 and TWA 27 A would be more evolved disks with a high C/O ratio. Lienert et al. (in prep) explore the case when the gap in the disk is due to internal photo evaporation and their model can reproduce a C/O >1 ratio as inferred in VLMS and BDs when using a reduced photo evaporative mass  loss rate, which lead to delayed gap opening times.

\subsection{H$_2$ emission}
\begin{figure}
   \centering
   \includegraphics[width=\hsize]{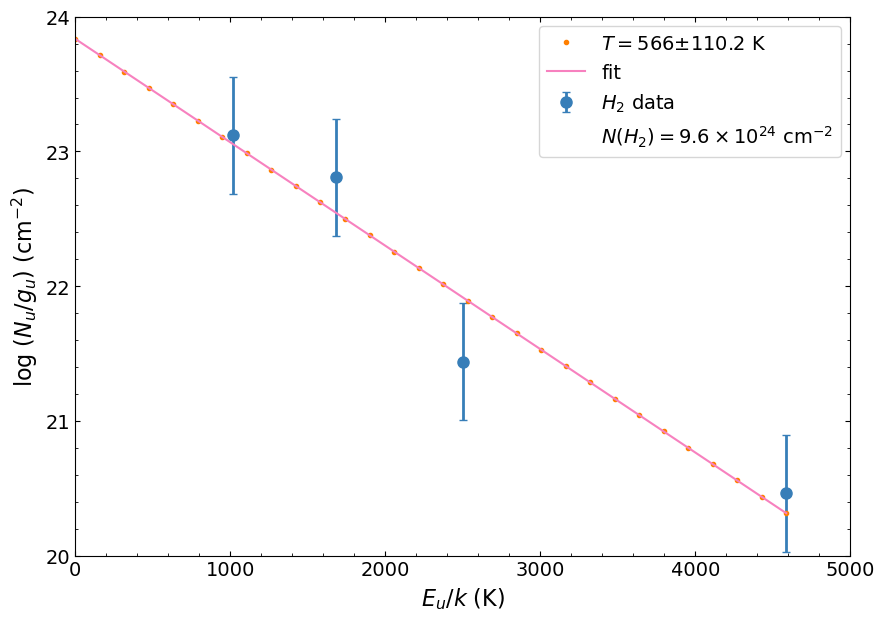}
      \caption{ Rotation diagram of H$_2$ pure rotational transitions. A linear fit performed on the estimated fluxes is shown  as a single component for the emitting gas with a temperature of about 570 K.}
               \label{Fig:rotD}
   \end{figure}

We can use the estimated fluxes to build a rotational diagram \citep{Goldsmith99} to estimate the emitting gas temperature assuming that the H$_2$ emission is in local thermodynamical equilibrium (LTE). Following \cite{Franceschi24} we obtain the rotation diagram presented  in Figure~\ref{Fig:rotD}. This figure has been produced using the tool pdrtpy, introduced in \cite{Berne22, Pound23}. Since the emitting radius of the H$_2$ emission cannot be directly estimated due to its optically thin nature, we assume the H$_2$ emission has the same emitting radius as the C$_2$H$_2$ optically thicker emission, 0.06 au. Using this estimate for the emitting radius, we found a temperature of 566~$\pm$~110 K. This temperature is consistent with the of H$_2$ emission coming from the same region of the disk as hydrocarbons. In addition, the emission is observed at a velocity of 20~$\pm$~6~km s$^{-1}$ consistent with the rest velocity of the star \cite[16.35~$\pm$~0.63, ][]{Joergens06}. However, due to the uncertainties of the line intensities because of the underlying emission of other molecules further investigation will have to await for a proper modeling of all molecules involved.
H$_2$ pure rotational lines are found to be relatively weak toward T Tauri stars with respect to emission from other molecules and  atoms. This emission was first observed in BD disks by \cite{Pascucci13} who detected either the S(1) or the S(2) transitions in half of their BD sample with Spitzer/IRS spectra. Among the BD disks with a MIRI spectrum, the three of them show  H$_2$ pure rotational emission. J0438 presents five pure rotational bands of H$_2$ with marginally extended emission for the H$_2$ S(1), S (3) and S (5) lines. The authors attribute this emission to a disk wind and/or to the disk surface. For TWA 27 A only H2 S(1) is clearly detected.

\section{Conclusions}\label{Conclusions}

We have used JWST to probe the inner, dense, and warm region of the disk around the BD \chaha1, where small planets may form. We have used the 5 to 26~$\mu$m spectrum to investigate the dust and gas emission and we draw the following conclusions:

\begin{itemize}

\item{\chaha1 presents a strong silicate feature. The dust is primarily composed of large ($\sim$4~$\mu$m) amorphous grains, indicating that grain growth has already begun by $\sim$1.5 Myr, consistent with previous studies. Amorphous SiO$_2$ is barely detected. For crystalline silicates, only large grains of forsterite are observed. Unlike other brown dwarf disks, there is no reliable detection of enstatite. No PAH features are found either.}

\item{The spectrum exposes numerous molecular features, with particularly strong emission of the organic molecule C$_2$H$_2$ at 13.7~$\mu$m. Additional molecular features include various hydrocarbons (C$_6$H$_6$, C$_3$H$_4$, C$_4$H$_2$, C$_2$H$_4$, C$_2$H$_6$, CH$_4$, and CH$_3$), as well as CO$_2$, HCN, H$_2$, and H$_2$O. Two isotopologues, $^{13}$CCH$_2$ and $^{13}$CO$_2$, have also been identified. The diversity of carbon-bearing molecules suggests a carbon-rich inner disk environment, as observed in the BD TWA 27 A \citep{Patapis25} and some VLMSs \citep{Tabone23,Arabhavi24, Kanwar24}. However, this is the only case where warm water is detected alongside all the hydrocarbons. We do not detect CO or OH.}

\item{The hydrocarbons exhibit temperatures between 225-450 K, suggesting they originate from a shared reservoir. They emit from a compact region within 0.1 au of the host star. In contrast, CO$_2$ appears to emit from farther out and/or deeper within the disk, while HCN is significantly warmer, likely indicating its emitting area is closer to the star or higher up in the disk.}

\item{Several H$_2$ pure rotational lines are clearly identified. The gas temperature and velocity point towards the emission most probably coming from the very inner, warm surface of the disk, the same region of the disk as hydrocarbons.}

\item{Based on models by \citet{Mah23} and \cite{Kanwar24} we hypothesize that \chaha1—with faint water signatures, various hydrocarbons, shallow molecular continuum on top of a strong silicate feature— could be transitioning from a very young disk rich in water to a more evolved disk with a high C/O ratio.}

\end{itemize}

The spectrum revealed by JWST/MIRI for \chaha1 shows the most diverse  chemistry seen to date in a BD protoplanetary disk, consisting of a strong dust feature, 12 carbon-bearing molecules plus H$_2$, and water.  The rich molecular environment presents a unique opportunity to test our understanding of disk chemistry and its influence on potential planet formation. Gaining insight into the reservoirs of the detected molecules can further illuminate the composition of future planets forming around brown dwarfs.

\begin{acknowledgements}
We are grateful to the referee, Dr. Ralf Siebenmorgen, for his useful comments and suggestions to the original manuscript. This work is based on observations made with the NASA/ESA/CSA James Webb Space Telescope. The data were obtained from the Mikulski Archive for Space Telescopes at the Space Telescope Science Institute, which is operated by the Association of Universities for Research in Astronomy, Inc., under NASA contract NAS 5-03127 for JWST. These observations are associated with program \#1282. The following National and International Funding Agencies funded and supported the MIRI development: NASA; ESA; Belgian Science Policy Office (BELSPO); Centre Nationale d’Etudes Spatiales (CNES); Danish National Space Centre; Deutsches Zentrum fur Luft- und Raumfahrt (DLR); Enterprise Ireland; Ministerio De Econom\'ia y Competividad; Netherlands Research School for Astronomy (NOVA); Netherlands Organisation for Scientific Research (NWO); Science and Technology Facilities Council; Swiss Space Office; Swedish National Space Agency; and UK Space Agency.

M.M.C. and D.B. have been funded by grants No. PID2019-107061GB-C61 and PID2023-150468NB-I00 by the Spain Ministry of Science, Innovation/State Agency of Research MCIN/AEI/ 10.13039/501100011033 and by “ERDF A way of making Europe.

A.C.G. acknowledges support from PRIN-MUR 2022 20228JPA3A “The path to star and planet formation in the JWST era (PATH)” funded by NextGeneration EU and by INAF-GoG 2022 “NIR-dark Accretion Outbursts in Massive Young stellar objects (NAOMY)” and Large Grant INAF 2022 “YSOs Outflows, Disks and Accretion: towards a global framework for the evolution of planet forming systems (YODA)”.

G.P. gratefully acknowledges support from the Carlsberg Foundation, grant CF23-0481 and from the Max Planck Society.

E.v.D. acknowledges support from the ERC grant 101019751 MOLDISK and the Danish National Research Foundation through the Center of Excellence ``InterCat'' (DNRF150).

T.H. and K.S. acknowledge support from the European Research Council under the Horizon 2020 Framework Program via the ERC Advanced Grant Origins 83 24 28.

I.K., A.M.A., and E.v.D. acknowledge support from grant TOP-1 614.001.751 from the Dutch Research Council (NWO).

I.K., and J.K. acknowledge funding from H2020-MSCA-ITN-2019, grant no. 860470 (CHAMELEON).

B.T. is a Laureate of the Paris Region fellowship program, which is supported by the Ile-de-France Region and has received funding under the Horizon 2020 innovation framework program and Marie Sklodowska-Curie grant agreement No. 945298.

O.A. and V.C. acknowledge funding from the Belgian F.R.S.-FNRS.

I.A., D.G. and B.V. thank the Belgian Federal Science Policy Office (BELSPO) for the provision of financial support in the framework of the PRODEX Programme of the European Space Agency (ESA).

L.C. acknowledges support by grant PIB2021-127718NB-I00,  from the Spanish Ministry of Science and Innovation/State Agency of Research MCIN/AEI/10.13039/501100011033.

T.P.R acknowledges support from ERC grant 743029 EASY.

D.R.L. acknowledges support from Science Foundation Ireland (grant number 21/PATH-S/9339).

M.T., M.V. and A.D.S acknowledge support from the ERC grant 101019751 MOLDISK.

VC thanks the Belgian F.R.S.-FNRS, and the Belgian Federal Science Policy Office (BELSPO) for the provision of financial support in the framework of the PRODEX Programme of the European Space Agency (ESA) under contract number 4000142531.

T.K. acknowledges support from STFC Grant ST/Y002415/1.
\end{acknowledgements}

\onecolumn
\begin{appendix}

\section{Continuum definition}\label{App:Continuum}

\begin{figure*}[ht]
   \centering
   \includegraphics[width=\hsize]{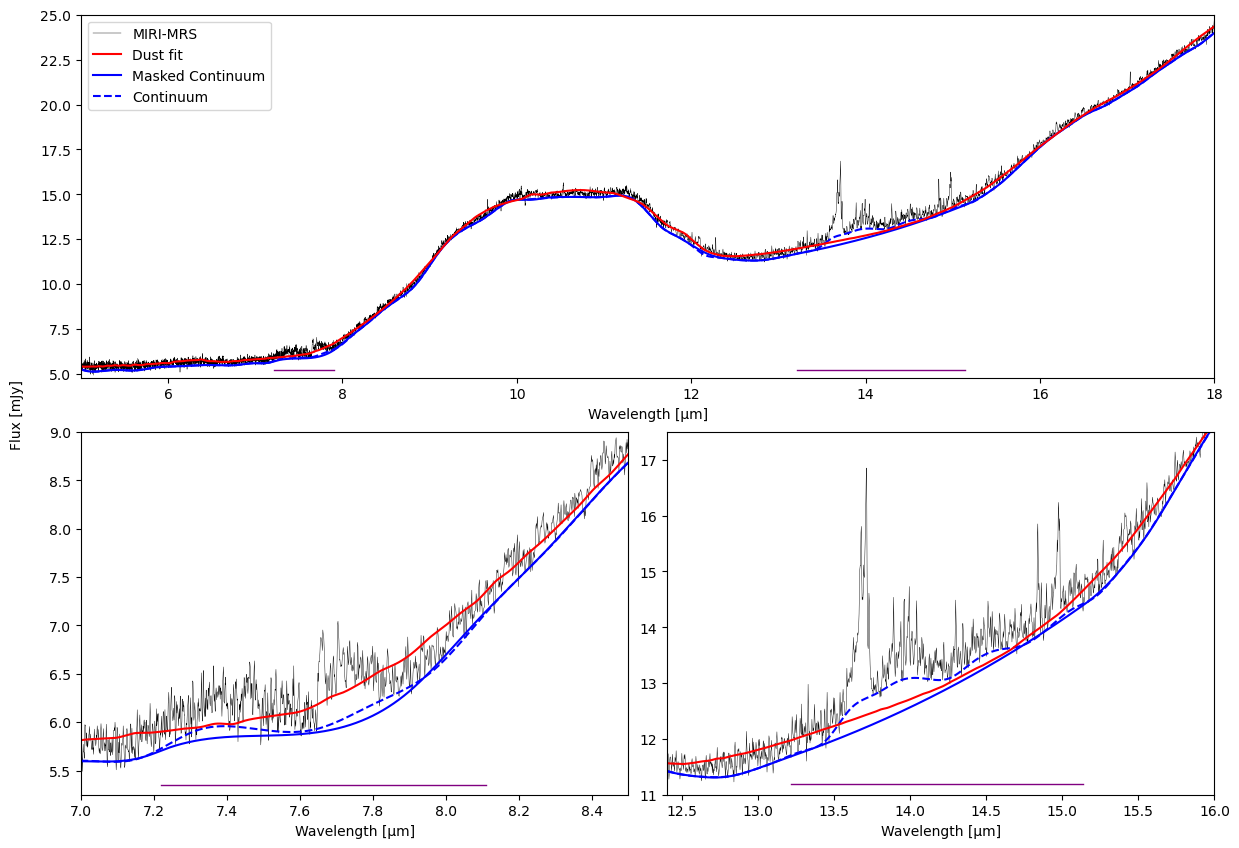}
      \caption{ Continuum determination is shown in the top panel. The lower panels are zooms of the regions around 7.7~\mum (left) and 14~\mum (right). The red line represents the dust fit performed in Section~3. The dashed blue line is the continuum resulting from using a small width Savitzky-Golay filter. The final masked continuum adopted in this work is the solid blue line while the windows masked are represented by horizontal lines at the bottom of each panel}
               \label{Fig:App:cont}
   \end{figure*}

For gas fitting using the 0D slab models, the continuum must be more accurately determined than what a dust fitting tool—constrained by its opacity and grain size assumptions—can provide. Therefore, we did not use the output from the dust retrieval code shown in Figure~\ref{Fig:dust} directly, as its continuum slightly overestimates the observed data at several wavelengths. Additionally, we suspected the presence of a molecular pseudo-continuum. To address this, we adopted a continuum determination procedure based on the method described in \cite{Temmink24}, ensuring a more reproducible and physically consistent continuum estimate. 
This procedure estimates the dust continuum emission using the pybaselines package \citep{Erb22}. Initially, a dust continuum level is estimated using a Savitzky-Golay filter with a third order polynomial. Then, all emission lines extending above 2$ \sigma$ of the standard deviation are masked until only the continuum remained. We explored different widths of the Savitzky-Golay filter. Larger widths would not trace well the area around 10~\mum, due the shape of the silicate emission bump. Thus we used a small width (blue dashed line in Figure~\ref{Fig:App:cont}), however subtracting the continuum defined with the narrow width subtracts also the molecular pseudo-continuum. In order to retain the molecular-pseudo-continuum we defined two windows around 7.7 and 14~\mum and use a cubic-spline function to interpolate between the ends of those windows. The resulting continuum is the blue line in Figure~\ref{Fig:App:cont}. 

\FloatBarrier

\section{Continuum test}\label{App:fits}

\begin{figure}[h!]
   \centering
   \includegraphics[width=10cm]{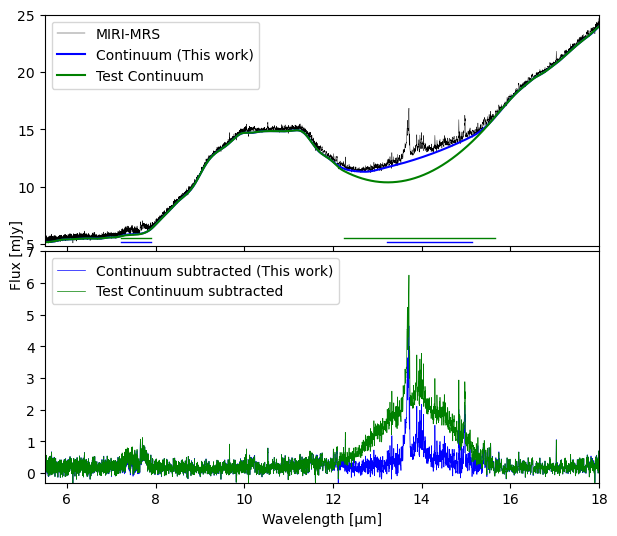}
      \caption{Continuum test. The upper panel shows the continuum used in this work (blue line) compared to the continuum defined with the largest window tested (green line). The windows used in each case are shown as horizontal lines. The lower panel shows the continuum subtracted spectra with the same code of colors. }
               \label{Fig:App:conttest}
   \end{figure}

\begin{figure}[h!]
   \centering
   \includegraphics[width=10cm]{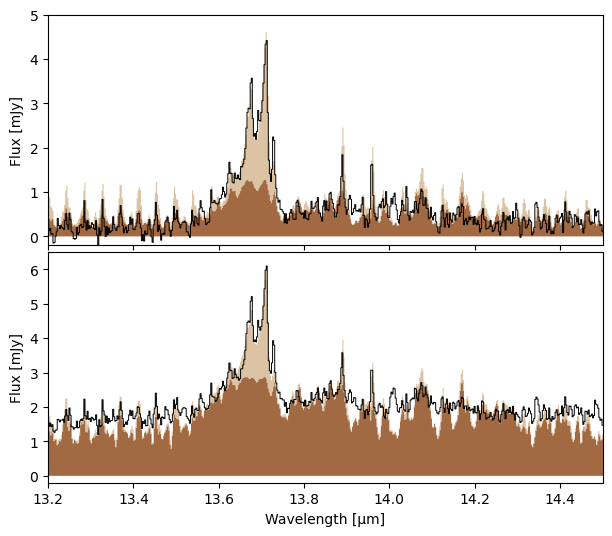}
      \caption{ Continuum test. Comparison of the C$_2$H$_2$ slab model result from the best fit used in this work (upper panel) and using the continuum shown in green in Fig.~\ref{Fig:App:cont} (lower panel). Both panels show a fit using two C$_2$H$_2$ slab models with higher (dark brown) and lower (light brown) column densities.  }
         \label{Fig:App:c2h2}
   \end{figure}

Because the continuum definition can be highly subjective when relying on a visual identification of line-free regions, we opted to choose a more reproducible approach following the method described in \cite{Temmink24}. Nevertheless, during the initial fitting, the presence of a molecular pseudo-continuum became evident, needing the use of the windowing technique outlined in Appendix~\ref{App:Continuum} to fit the spectral features. However, the definition of this window still involves a degree of subjectivity. In order to investigate the impact in our results when using different continuum definitions we explored several window sizes in the spectral region around 14~\mum. The results corresponding to the largest window tested are presented in Figure~\ref{Fig:App:conttest}, which shows both the continuum and continuum subtracted spectra, and in Figure~\ref{Fig:App:c2h2}, which displays the fit of C$_2$H$_2$. We present the results obtained for the fitting with the tested continuum in Table~\ref{Tab:App:continuumtest} in order to compare them with the parameters obtained in this work. As a result from this test we find that the same molecules are detected, or tentatively detected, in all cases and the maximum change in the parameters is 100~K and 1 order of magnitude in temperatures and column densities respectively. 

\begin{table*}[h!]
\caption{Comparison of parameters derived in this work with the ones obtained from the continuum test.}    
\label{Tab:App:continuumtest}
\centering                         
\begin{tabular}{l |c c c | c c c}
\hline\hline                 
Species & $T[K]$ & N[10$^{18}$cm$^{-2}$] & R[au] & $T[K]$ & N[10$^{18}$cm$^{-2}$] & R[au] \\  
\hline                        
              &\multicolumn{3}{c|}{This work} &  \multicolumn{3}{c}{Continuum test} \\
\hline
C$_2$H$_2$ + $^{13}$CCH$_2$ & 300 & 316 & 0.06 & 350 & 681& 0.04 \\    

C$_2$H$_2$ + $^{13}$CCH$_2$ & 300 & 0.7 & 0.07 & 350 & 0.5 & 0.06 \\  

HCN & 525 & 1   & 0.02 & 500 & 1 & 0.03  \\

CO$_2$ +  $^{13}$CO$_2$ & 175 & 68     & 0.15 & 225 & 2 & 0.11\\

C$_6$H$_6$ & 225 &0.07\tablefootmark{b}    & 0.09 & 275 & 0.1\tablefootmark{b} & 0.06  \\

C$_3$H$_4$ & \multicolumn{3}{c|}{Tentative Detection}  & \multicolumn{3}{c}{Tentative Detection}\\ 

C$_4$H$_2$ &  \multicolumn{3}{c|}{Detected}  &  \multicolumn{3}{c}{Detected}  \\

C$_2$H$_4$ &  \multicolumn{3}{c|}{Tentative Detection}  & \multicolumn{3}{c}{Tentative Detection}\\

C$_2$H$_6$ &  \multicolumn{3}{c|}{Tentative Detection}  &  \multicolumn{3}{c}{Tentative Detection}\\

CH$_4$ & 450 & 100 & 0.02  & 350 & 46 & 0.06\\      

H$_2$O   &  \multicolumn{3}{c|}{Detected}  &  \multicolumn{3}{c}{Detected} \\

CH$_3$ &  \multicolumn{3}{c|}{Detected}   &  \multicolumn{3}{c}{Detected}   \\

\hline                               
\end{tabular}
\tablefoot{
\tablefoottext{b}{Parameter not well constrained.} 
}
\end{table*}

\FloatBarrier

\section{Molecules with no quantitative information}\label{App:tentatives}

For some molecules, when slab model fits were attempted, the remaining degeneracies in column density and temperature were too significant to allow for a meaningful analysis. Those molecules are labeled as "Detected" or "Tentative detection" in Table~\ref{Tab:molecules}. We provide in Table~\ref{Tab:App:molecules} the parameters of the slab models used to plot the emission of these molecules in this work. Note that in some cases, e.g. water, the degeneracy is very large (see the $\chi^2$ maps in Fig.~\ref{Fig:xis}). In the case of CH$_3$ we did not fit the molecule due to limitations defined in Sect.~\ref{Results:Gas}.

\begin{table}[h!]
\caption{Temperatures, column densities and emitting radii of the slab models used to show the ‘detected’ and ‘tentatively detected’ molecules in the spectrum of \chaha1.}           
\label{Tab:App:molecules}     
\centering                         
\begin{tabular}{l c c c}      
\hline\hline                 
Species & $T[K]$ & N[cm$^{-2}$] & R[au] \\ 
\hline                        

C$_3$H$_4$ & 250 & 4.6$\times$10$^{14}$ & 0.10 \\ 

C$_4$H$_2$ & 225 & 1.5$\times$10$^{14}$ & 0.11  \\

C$_2$H$_4$ & 200 & 1.5$\times$10$^{19}$ & 0.09  \\   

C$_2$H$_6$ &  350 & 4.6$\times$10$^{19}$& 0.02  \\

H$_2$O   & 550 & 1$\times$10$^{22}$ & 0.01   \\

CH$_3$ &   400 & 2.2$\times$10$^{16}$ & 0.09   \\

\hline                        
\end{tabular}

\end{table}
\FloatBarrier

\section{H$_2$ emission}\label{App:H2emission}
\begin{figure}
   \centering
   \includegraphics[width=10cm]{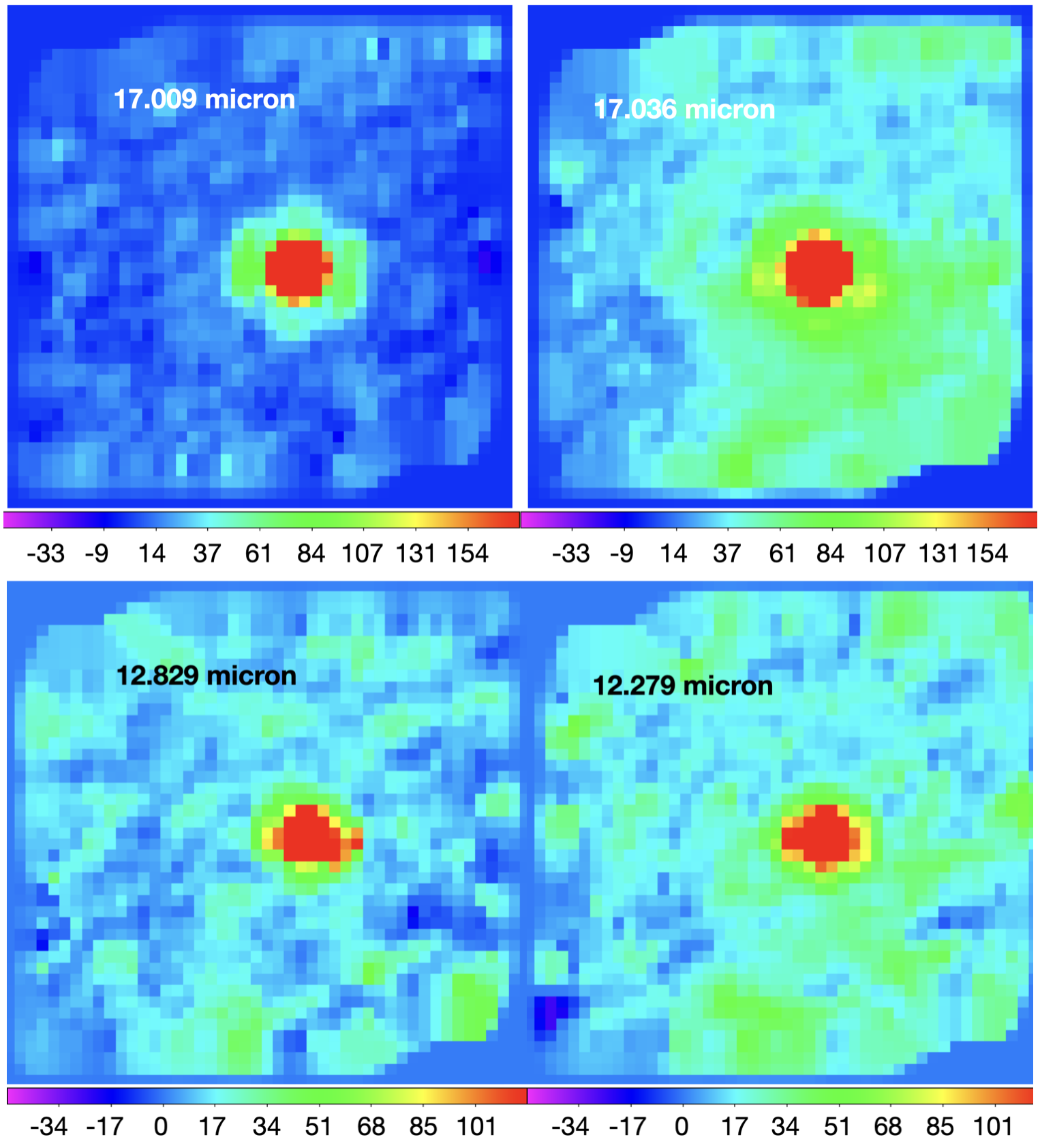}
      \caption{IFU aligned images of the H$_2$ S(1) line. The panels on the left show wavelengths with no emission at 17.009~\mum (top) and 12.829~\mum (bottom).The right panels show  the emission at the line rest-wavelengths for H$_2$ S(1) at the top and H$_2$ S(2) at the bottom. }
         \label{Fig:App:h2extended}
   \end{figure}

The emission of the S(1) and S(2) H$_2$ lines may be somewhat extended. In Fig.~\ref{Fig:App:h2extended} we show IFU aligned images of the H$_2$ S(1) line. The panel on the left shows a wavelength with no emission, the central panel shows  the emission at the  H$_2$ S(1) line wavelength and in the right panel we see the difference between both images. A structure is hinted in the lower right part of the image.
Another indication that the structure may be real is that these lines appear much stronger (while all other lines in the spectrum stay the same) if we use a background subtraction method such as sdither. In this case we leverage the four-point dither pattern to obtain a first guess on the background map, then refine it using a median filter which both smooths the background estimate and removes residual star signals from it. When using our preferred background subtraction method,  ddither, we do a direct pairwise dither subtraction and thus, we could be removing part of this emission making the line in the spectrum to show fainter. 

\end{appendix}

\end{document}